\documentclass[journal,10pt]{IEEEtran}
\usepackage{amsmath,epsfig,subfigure,cite,amssymb,graphicx,amsfonts,color,bm,multirow,stfloats,float,tabu,tabularx,bbm}
\usepackage{algorithm}
\usepackage{algorithmicx}
\usepackage{algpseudocode}
\usepackage{booktabs}
\usepackage{mathrsfs}

\usepackage{epstopdf}
\usepackage{enumerate}
\usepackage{balance}

\begin{document}
\title{Fast Direct Localization for Millimeter Wave MIMO Systems via Deep ADMM Unfolding}
\author{Wenzhe Fan, \emph{Graduate Student Member, IEEE}, Shengheng Liu, \emph{Senior Member, IEEE},\\ Chunguo Li, \emph{Senior Member, IEEE}, and Yongming Huang, \emph{Senior Member, IEEE}
	
\thanks{This work was supported in part by the National Natural Science Foundation of China under Grant Nos. U1936201 and 62001103. (\emph{Corresponding authors: Yongming Huang and Shengheng Liu})}
\thanks{The authors are with the School of Information Science and Engineering, Southeast University, Nanjing 210096, China, and also with the Purple Mountain Laboratories, Nanjing 211111, China  (email: \{wzfan, s.liu, chunguoli, huangym\}@seu.edu.cn). }
\vspace{-2em}
}

\maketitle

\begin{abstract}
Massive arrays deployed in millimeter-wave systems enable high angular resolution performance, which in turn facilitates sub-meter localization services. Albeit suboptimal, up to now the most popular localization approach has been based on a so-called two-step procedure, where triangulation is applied upon aggregation of the angle-of-arrival (AoA) measurements from the collaborative base stations. This is mainly due to the prohibitive computational cost of the existing direct localization approaches in large-scale systems. To address this issue, we propose a deep unfolding based fast direct localization solver. First, the direct localization is formulated as a joint $l_1$-$l_{2,1}$ norm sparse recovery problem, which is then solved by using alternating direction method of multipliers (ADMM). Next, we develop a deep ADMM unfolding network (DAUN) to learn the ADMM parameter settings from the training data and a position refinement algorithm is proposed for DAUN. Finally, simulation results showcase the superiority of the proposed DAUN over the baseline solvers in terms of better localization accuracy, faster convergence and significantly lower computational complexity.

\end{abstract}
\IEEEpeerreviewmaketitle
\vspace{-0.3cm}
\section{Introduction}
\IEEEPARstart{M}{illimeter-wave} (mmWave) multiple-input-multiple-output (MIMO) has been recognized as the promising technologies in fifth/sixth generation cellular communication networks to cope with the increasing demand in data rate and system capacity, thanks to the high spectral efficiency and beam directivity \cite{xiao2017millimeter}. Aside from the unprecedented transmission capacity, investigations on mmWave MIMO system also has shown the potential of achieving sub-meter user localization and the related value-added services \cite{shahmansoori2017position}, due to the high angular resolution provided by the large-scale arrays.

In the classic localization system, the user positioning usually comprises two steps,  which are first estimating the position-related parameters, e.g., AoA and time-of-arrival (ToA), and then recovering the user position from those parameters by trilateration or triangulation \cite{pan2022efficient}. In \cite{hu2014esprit}, the authors proposed an AoA-based approach for the 2-D localization in the massive MIMO systems. For the 3-D localization, a closed-form estimator was developed in \cite{wang2015asymptotically}. More recently, several novel localization methods for the indoor/outdoor mmWave MIMO systems were developed in \cite{shahmansoori2017position,yang2021model}.
However, the above two-step localization methods may yield poor performance for discarding useful information during the parameter estimation phase, because the essence of the estimated parameters is low-order moments of received signals \cite{zhao2020beamspace}.

As a more efficient alternative, the direct localization approach was proposed in \cite{weiss2004direct}, in which the user location was directly recovered from the received data without estimating intermediate parameters. In \cite{garcia2017direct}, a grid-based direct localization method was proposed for massive MIMO system, where the AoA-related user position was estimated by solving the convex sparse recovery problem. In \cite{lian2019user}, the user movement was considered in the direct localization system and an efficient variational Bayes solver was developed. Although the direct localization has great potential to improve the
localization performance, the most urgent issue imposed by current direct localization methods, e.g., \cite{garcia2017direct,zhao2020beamspace,lian2019user,zheng2020joint,li2022self}, is the heavy complexity, which is especially prominent and prohibits the practical implementation in the mmWave MIMO systems.

To address the issue mentioned above, we propose a fast direct localization method for mmWave MIMO systems. The direct localization is first formulated as a joint $l_1$-$l_{2,1}$ norm sparse recovery problem. Note that no existing compressed sensing methods, e.g., iterative shrinkage thresholding algorithm (ISTA) \cite{chambolle1998nonlinear}, approximate message passing (AMP) \cite{rangan2019convergence} or sparse Bayesian learning (SBL) \cite{tzikas2008variational}, are applicable for such joint $l_1$-$l_{2,1}$ norm sparse recovery problem. Our main contributions are: 1) we develop a dedicated alternating direction method of multipliers (ADMM) solver for direct localization with quite low complexity; 2) we unroll the ADMM solver into a model-based deep network to learn a better iterative performance of ADMM; 3) we design a position refinement algorithm to further improve the accuracy of the proposed deep ADMM unfolding network (DAUN). Overall, the proposed method is a first attempt to tackle such $l_1$-$l_{2,1}$ norm direct localization problem by ADMM and unfolding architecture. The simulations indicate that the DAUN has higher localization accuracy, faster convergence speed, and much lower computational complexity, as compared with other baselines.

\textit{Notations: }  Lower (upper)-case bold characters are used to denote vectors (matrices), and the vectors are by default in column orientation. The superscripts $(\cdot)^{\mathsf T}$ and and $(\cdot)^{\mathsf H}$ represent the transpose and conjugate transpose operators, respectively. $\mathbb{E}$ returns the expected value of a discrete random variable.
${\left\|  \cdot  \right\|_1}$, ${\left\|  \cdot  \right\|_2}$ and ${\left\|  \cdot  \right\|_{2,1}}$ denote the $l_1$ norm, $l_2$ and $l_{2,1}$ norms, respectively.
The operator ${\rm{vec(}} {\bf X} {\rm{)}}$ vectorizes a matrix ${\bf X}$, and ${\rm{blkdiag(}}{{\bf{X}}_1}{\rm{,}}{{\bf{X}}_2}, \ldots ,{{\bf{X}}_M}{\rm{)}}$ denotes the block diagonal matrix with  ${{\bf{X}}_1}{\rm{,}}{{\bf{X}}_2}, \ldots ,{{\bf{X}}_M}$ being in its diagonal positions.

\section{Preliminaries} \label{sec:sys}
\subsection{Signal Model}

We consider a two-dimensional target area $\mathcal{R}$ with a single-antenna user and $M$ base stations (BSs) equipped with the array of $N_m$ antennas each.
The centers of the gravity of arrays are located at ${{{\bf{\tilde p}}}_m} = [\tilde p_m^x,\tilde p_m^y], m=1,2,\ldots,M$, which are precisely known and assumed to be in the far field with respect to the user. The user location is ${{\bf{p}}} = [p^x,p^y] \in \mathcal{R}$ and it broadcasts the pilot signal ${s_{\rm{d}}}$ sounded by all the BSs. The received narrowband signal at the $m$-th BS is given as
\begin{eqnarray}
	{{\bf{y}}_m} = \sqrt \omega  {{\bf{h}}_m}{s_{\rm{d}}} + {{\bf{n}}_m},m = 1,2, \ldots ,M,
\end{eqnarray}
where $\omega$ is the signal-to-noise ratio (SNR), ${{\bf{h}}_{m}}$ is the mmWave channel vector between the user and the $m$-th BS, and  ${{\bf{n}}_m} \sim {\mathcal{CN}}({\bf{0}},{{\bf{I}}_{{N_m}}})$ is the received noise with normalized power. Typically, the mmWave channel vector ${{\bf{h}}_m}$ is composed of one line of sight (LoS) path and several non-LoS (NLoS) ones, which can be expressed as ${{\bf{h}}_m} = {\alpha _m}{{\bf{a}}_m}({\theta _m}({\bf{p}})) + \sum\nolimits_{i = 1}^{{P_m}} {\alpha _m^i{{\bf{a}}_m}(\theta _m^i)} $,
where ${{P_{m}}}$ is the number of NLoS paths, ${\bf{a}}_{m}(\theta )$ is the array response, ${{\alpha _{m}}}$ and ${\theta _{m}}({{\bf{p}}})$ respectively denote the path gain and the angle of arrival (AoA) of LoS path, $\{ {\alpha _{m}^i}\} _{i = 1}^{{P_{m}}}$ and $\{ {\theta _{m}^i}\} _{i = 1}^{{P_{m}}}$ respectively denote the path gains and the AoAs of NLoS paths.
The AoA of LoS path is associated with the user position through ${\theta _{m}}({{\bf{p}}}) = \arctan (({p^y} - \tilde p_m^y)/({p^x} - \tilde p_m^x))$. Finally, the received signals of all the BSs are sent to the fusion center for joint processing\footnote{The proposed localization scheme can be extended to the self-position awareness systems, in which the BSs (anchors) transmit the signals consisting of the physical positions of BSs (anchors) and pilot sequence for channel information acquisition. The user equipment fuses all the received signals and recover its position without uplink transmission.}.

\vspace{-0.3cm}
\subsection{Problem Formulation }
Since the AoAs of user are sparse in angular domain and the position is an AoA-related parameter, we can exploit compressed sensing method to
directly estimate ${{\bf{p}}}$ from the received signal ${{\bf{y}}_m}$.
The main idea is to sample the continuous map into grids and find a sparse representation of the mmWave channel ${{\bf{h}}_{m}}$. In this way, the target area $\mathcal{R}$ can be sampled into $K$ uniform grid locations, which is given by
${\mathcal{K}} = \{ {{\bf{\Phi }}_1},{{\bf{\Phi }}_2}, \ldots ,{{\bf{\Phi }}_K}\} \in \mathcal{R}$,
where ${{\bf{\Phi }}_k} = [\phi _k^x,\phi _k^y]$ is the coordinate of the $k$-th grid location. Then, we further define a uniform grid of $L_m$ angles, given by ${{\mathcal{L}}_m} = \{ {\vartheta _1},{\vartheta _2}, \ldots ,{\vartheta _{{L_m}}}\}$, where $L_m \gg P_m,m=1,2,\ldots,M$. If the above grid area is densely discretized, i.e., $K$ and ${{L_m}}$ are sufficiently large, the received signal ${{\bf{y}}_m}$ can be expressed as
\begin{eqnarray}
	{{\bf{y}}_m} = {{\bf{A}}_m}{{\bf{x}}_m} + {\bf{B}}_m{{\bf{z}}_m} + {{\bf{n}}_m},m = 1,2, \ldots ,M,
\end{eqnarray}
where ${{\bf{A}}_m} = [{{\bf{a}}_m}({\theta _m}({{\bf{\Phi }}_1})),{{\bf{a}}_m}({\theta _m}({{\bf{\Phi }}_2})), \ldots ,{{\bf{a}}_m}({\theta _m}({{\bf{\Phi }}_K}))]$, ${{\bf{B}}_m} = [{{\bf{a}}_m}({\vartheta _1}),{{\bf{a}}_m}({\vartheta _2}), \ldots ,{{\bf{a}}_m}({\vartheta _{{L_m}}})]$, ${\theta _m}({{\bf{\Phi }}_k})$ is the AoA from the $k$-th
grid to the $m$-th BS, ${{\bf{x}}_m} = {[{x_{m,1}},{x_{m,2}}, \ldots ,{x_{m,K}}]^T}$ is a sparse vector with ${x_{m,l}}$ corresponding to the effective gain of the LoS path from the grid location $k$ to $m$-th BS and there exists only one non-zero element in ${{\bf{x}}_m}$. Meanwhile, ${{\bf{z}}_m} = {[{z_{m,1}},{z_{m,2}}, \ldots ,{z_{m,L_m}}]^{\mathsf T}}$ is also a sparse vector with ${z_{m,l}}$ corresponding to the effective gain of the NLoS path from the user to $m$-th BS with the grid angle ${\vartheta _l}$. We next define a matrix ${\bf{X}} = [{{\bf{x}}_1},{{\bf{x}}_2}, \ldots ,{{\bf{x}}_M}]$. Since only one element in ${{\bf{x}}_m}$ is non-zero and it corresponds to the user location, the matrix ${\bf{X}}$ is row-sparse.  Based upon this property, the direct localization problem is formulated as
\vspace{-0.2cm}
\begin{eqnarray}
	\begin{array}{*{20}{c}}
	\mathop {\min }\limits_{{\bf{\hat X}},\{ {{{\bf{\hat z}}}_m}\} _{m = 1}^M} &{{{\left\| { \bf{\hat X}} \right\|}_{2,1}} + \sum\limits_{m=1}^M {{w_m}{{\left\| {{{\bf{\hat z}}_m}} \right\|}_1}} }\\
	{{\rm{s}}{\rm{.t}}.}&{{{\bf{y}}_m} = {{\bf{A}}_m}{{\bf{\hat x}}_m} + {\bf{B}}_m{{\bf{\hat z}}_m},\forall m.}
	\end{array}
	\label{eq:direct}
\end{eqnarray}
where ${ \bf{\hat X}}$ and ${\bf{\hat z}}_m$ respectively denote the estimates of ${ \bf{ X}}$ and ${\bf{ z}}_m$, and $\{ {w_m}\} _{m = 1}^M$ is an unknown weighted coefficient.
After solving \eqref{eq:direct}, the position estimate is given by ${{\bf{\Phi }}_{\hat k}}$, where the index $\hat k = \mathop {\arg \max }_k {\left\| {[{{\bf{\hat X}}]_{k,:}}} \right\|_2}$ is the active row in ${\bf{\hat X}}$.

\textit{Remark 1:} Notice that although the problem \eqref{eq:direct} is convex, the complexities of effective convex solvers are cubic with respect to the grid numbers $K$ and $L_m$, which is computationally intractable for real-time localization in the mmWave MIMO systems. To address this issue, a deep ADMM unfolding framework for  direct localization is designed in next section.

\vspace{-0.2cm}
\section{Algorithm Description}

\subsection{ADMM Solver for Direct Localization}
\vspace{-0.1cm}
In essence, the direct localization problem \eqref{eq:direct} is a joint $l_1$-$l_{2,1}$ norm optimization but there are no dedicated compressed sensing methods to tackle such optimization. According to the form of problem \eqref{eq:direct}, the global optimum can be attained by using the ADMM. The first step of ADMM is to give the augmented Lagrangian function of problem \eqref{eq:direct}, which is
\begin{eqnarray}
	{J_\rho }({\bf{\hat X}},{\bf{\hat z}},{\bf{s}}) \!\!\!\!\!&=&\!\!\!\!\! {\left\| {\bf{\hat X}} \right\|_{2,1}} + \sum\limits_m^M {{w_m}{{\left\| {{{\bf{\hat z}}_m}} \right\|}_1}}  + {{\bf{s}}^{\mathsf H}}({\bf{A \hat x}} + {\bf{B \hat z}} - {\bf{y}}) \nonumber\\
	\!\!\!\!\!&+&\!\!\!\!\! \frac{\rho }{2}\left\| {{\bf{A \hat x}} + {\bf{B \hat z}} - {\bf{y}}} \right\|_2^2,
\end{eqnarray}
where ${\bf{A}} = {\rm{blkdiag(}}{{\bf{A}}_1},{{\bf{A}}_2}, \ldots ,{{\bf{A}}_M}{\rm{)}}$, ${\bf{\hat x}} = {\rm{vec(}}{\bf{\hat X}}{\rm{)}}$, ${\bf{B}} = {\rm{blkdiag(}}{{\bf{B}}_1},{{\bf{B}}_2}, \ldots ,{{\bf{B}}_M}{\rm{)}}$, ${\bf{\hat z}} = [{\bf{\hat z}}_1^{\mathsf T},{\bf{\hat z}}_2^{\mathsf T}, \ldots ,{\bf{\hat z}}_M^{\mathsf T}]^{\mathsf T}$, ${\bf{y}} = {[{\bf{y}}_1^{\mathsf T},{\bf{y}}_2^{\mathsf T}, \ldots ,{\bf{y}}_M^{\mathsf T}]^{\mathsf T}}$, ${\bf{s}} \in {{\mathbb{C}}^{\sum\nolimits_{m = 1}^M {{N_m}}  \times 1}}$ is the Lagrangian multiplier, and $\rho$ is a positive penalty parameter. Then the ADMM repeats the following calculations
\begin{eqnarray}
	{{\bf{\hat X}}^{(i + 1)}} = \mathop {\arg \min }\limits_{\bf{\hat X}} {J_\rho }({\bf{\hat X}},{{\bf{\hat z}}^{(i)}},{{\bf{s}}^{(i)}}), \label{eq:minX}\\
	{{\bf{\hat z}}^{(i + 1)}} = \mathop {\arg \min }\limits_{\bf{\hat z}} {J_\rho }({{\bf{\hat X}}^{(i + 1)}},{\bf{\hat z}},{{\bf{s}}^{(i)}}), \label{eq:minz}\\
	{{\bf{s}}^{(i + 1)}} = {{\bf{s}}^{(i)}} + \rho ({\bf{A\hat x}}^{(i+1)} + {\bf{B \hat z}}^{(i+1)} - {\bf{y}}),\label{eq:minS}	
\end{eqnarray}
until convergence. Now we focus on the problem \eqref{eq:minX}, the minimization is equivalent to
\vspace{-0.3cm}
\begin{eqnarray}
	\mathop {\min }\limits_{\bf{\hat X}} {\left\| {\bf{\hat X}} \right\|_{2,1}} + \frac{\rho }{2}\left\| {{\bf{A \hat x}} + {\bf{B}}{{\bf{\hat z}}^{(i)}} - {\bf{y}} + \frac{1}{\rho }{\bf{s}}^{(i)}} \right\|_2^2.
	\label{eq:second}
\end{eqnarray}

Then we approximate the second-order term in the objective function of \eqref{eq:second} by Taylor expansion at ${{{\bf{\hat x}}^{(i)}}}$ up to second order,
\vspace{-0.3cm}
\begin{eqnarray}
	\!\!\!\!\!&\left\| {{\bf{A\hat x}} + {\bf{B}}{{\bf{\hat z}}^{(i)}} - {\bf{y}} + \frac{1}{\rho }{{\bf{s}}^{(i)}}} \right\|_2^2&\!\!\!\!\! \!\approx\! \left\| {{\bf{A}}{{\bf{\hat x}}^{(i)}} \!+\! {\bf{B}}{{\bf{\hat z}}^{(i)}} \!-\! {\bf{y}} \!+\! \frac{1}{\rho }{{\bf{s}}^{(i)}}} \right\|_2^2 \nonumber\\
	&+&\!\!\!\!\!\!\!\!\!\!\!\!\!\!\!\!\!\!\!\!\!\!\!\!\!\!\!\!\!\!\!\!\!\!\!\!\!\! 2{({{\bf{g}}_1^{(i)}})^{\mathsf H}}({\bf{\hat x}} - {{\bf{\hat x}}^{(i)}}) + \frac{1}{{{\tau _1}}}\left\| {{\bf{\hat x}} - {{\bf{\hat x}}^{(i)}}} \right\|_2^2,
	\label{eq:tay}
\end{eqnarray}
where ${{\bf{g}}_1^{(i)}} = {{\bf{A}}^{\mathsf H}}({\bf{A}}{{\bf{\hat x}}^{(i)}} + {\bf{B}}{{\bf{\hat z}}^{(i)}} - {\bf{y}} + \frac{1}{\rho }{{\bf{s}}^{(i)}})$, and ${{\tau _1}}$ is a positive proximal parameter. Upon substituting \eqref{eq:tay} into \eqref{eq:second}, the minimization in \eqref{eq:second} becomes
\begin{eqnarray}
	\mathop {\min }\limits_{\bf{\hat X}} {\left\| {\bf{\hat X}} \right\|_{2,1}} + \rho {({{\bf{g}}_1^{(i)}})^{\mathsf H}}({\bf{\hat x}} - {{\bf{\hat x}}^{(i)}}) + \frac{\rho }{{2{\tau _1}}}\left\| {{\bf{\hat x}} - {{\bf{\hat x}}^{(i)}}} \right\|_2^2.
	\label{eq:min1}
\end{eqnarray}

After some mathematical manipulations on \eqref{eq:min1}, the minimization can be further expressed as
\begin{eqnarray}
	\mathop {\min }\limits_{\bf{\hat X}} {\lambda _1}{\left\| {\bf{\hat X}} \right\|_{2,1}} + \frac{1}{2}\left\| {{\bf{\hat x}} - {{\bf{c}}^{(i)}}} \right\|_2^2,
	\label{eq:min3}
\end{eqnarray}
where ${\lambda _1} = {\tau _1}/\rho $ and ${{\bf{c}}^{(i)}} = {{\bf{\hat x}}^{(i)}} - {\tau _1}{{\bf{g}}_1^{(i)}}$.

$\textit{Lemma 1:}$ The closed-form solution of the minimization in \eqref{eq:min3} can be expressed as
\begin{eqnarray}
	{[{\bf{\hat X}}^{(i+1)}]_{k,:}} \!=\!\! \left\{ {\begin{array}{*{20}{c}}
		\!\!\!\!{\frac{{{{\left\| {{{[{{\bf{C}}^{(i)}}]}_{k,:}}} \right\|}_2} - {\lambda _1}}}{{{{\left\| {{{[{{\bf{C}}^{(i)}}]}_{k,:}}} \right\|}_2}}}{{[{{\bf{C}}^{(i)}}]}_{k,:}},}&\!\!\!\!\!{{{\left\| {{{[{{\bf{C}}^{(i)}}]}_{k,:}}} \right\|}_2} \!>\! {\lambda _1}}\\
		\!\!\!\!{{{\bf{0}}^T},}&\!\!\!\!\!{{{\left\| {{{[{{\bf{C}}^{(i)}}]}_{k,:}}} \right\|}_2} \!\le\! {\lambda _1}}
		\end{array}} \right.
	\label{eq:solu}
\end{eqnarray}
where ${{\bf{C}}^{(i)}} = [{[{{\bf{c}}^{(i)}}]_{1:K}},{[{{\bf{c}}^{(i)}}]_{K + 1:2K}}, \ldots ,{[{{\bf{c}}^{(i)}}]_{(M - 1)K + 1:MK}}]$ and $k = 1, \ldots ,K$ .

\textit{Proof: } The problem \eqref{eq:min3} can be factorized into $M$ independent sub-problems on the rows of ${{\bf{\hat X}}}$, which is given by
\begin{eqnarray}
	\mathop {\min }\limits_{{{[{\bf{\hat X}}]}_{k,:}}} {\lambda _1}{\left\| {{{[{\bf{\hat X}}]}_{k,:}}} \right\|_2} + \frac{1}{2}\left\| {{{[{\bf{\hat X}}]}_{k,:}} - {{[{{\bf{C}}^{(i)}}]}_{k,:}}} \right\|_2^2,k = 1, \ldots ,K.
	\label{eq:chan}
\end{eqnarray}
By calculating the derivative with respect to ${{{[{\bf{\hat X}}]}_{k,:}}}$ and setting the it to zero, we have ${[{\bf{\hat X}}]_{k,:}} = 2{a_k}/({a_k} - {\lambda _1}){[{{\bf{C}}^{(i)}}]_{k,:}}$, where ${a_k} = {\left\| {{{[{\bf{\hat X}}]}_{k,:}}} \right\|_2}$. This implies that the solution to ${[{\bf{\hat X}}]_{k,:}}$ is linear with respect to ${[{{\bf{C}}^{(i)}}]_{k,:}}$ scaled by a scalar ${\eta _k} = 2{a_k}/({a_k} - {\lambda _1})$, i.e., ${[{\bf{\hat X}}]_{k,:}} = {\eta _k}{[{{\bf{C}}^{(i)}}]_{k,:}}$. In this way, the problem \eqref{eq:chan} becomes
\begin{eqnarray}
	\mathop {\min }\limits_{{{[{\bf{\hat X}}]}_{k,:}}} \!\!\!\!\!\!\!&\;&\!\!\!\!\!\!\! \frac{1}{2} \left\| {{{[{{\bf{C}}^{(i)}}]}_{k,:}}} \right\|_2^2\eta _k^2 + ({\lambda _1}{\left\| {{{[{{\bf{C}}^{(i)}}]}_{k,:}}} \right\|_2} - \left\| {{{[{{\bf{C}}^{(i)}}]}_{k,:}}} \right\|_2^2){\eta _k} \nonumber \\
	\!\!\!\!\!&+&\!\!\!\!\! \frac{1}{2}\left\| {{{[{{\bf{C}}^{(i)}}]}_{k,:}}} \right\|_2^2,
\end{eqnarray}
which is a quadratic function with respect to scalar ${\eta _k}$. By calculating the axis of symmetry of above quadratic function, the closed-form solution to ${\eta _k}$ is immediately obtained,  which is ${\eta _k} = ({\left\| {{{[{{\bf{C}}^{(i)}}]}_{k,:}}} \right\|_2} - {\lambda _1})/{\left\| {{{[{{\bf{C}}^{(i)}}]}_{k,:}}} \right\|_2}$ if ${\left\| {{{[{{\bf{C}}^{(i)}}]}_{k,:}}} \right\|_2} > {\lambda _1}$, and ${\eta _k} = 0$ if ${\left\| {{{[{{\bf{C}}^{(i)}}]}_{k,:}}} \right\|_2} \le {\lambda _1}$.  Then we arrive at the result given in \eqref{eq:solu}. \hfill $\blacksquare$

Next we focus on \eqref{eq:minz}. By performing similar derivations as \eqref{eq:second} and \eqref{eq:tay}, the minimization in \eqref{eq:minz} is equivalent to
\begin{eqnarray}
	\mathop {\min }\limits_{\bf{\hat z}} \sum\limits_m^M {{w_m}{{\left\| {{{\bf{\hat z}}_m}} \right\|}_1}}  + \rho {({\bf{g}}_2^{(i)})^{\mathsf H}}({\bf{\hat z}} - {{\bf{\hat z}}^{(i)}}) \!+ \! \frac{\rho }{{2{\tau _2}}}\left\| {{\bf{\hat z}} - {{\bf{\hat z}}^{(i)}}} \right\|_2^2,
	\label{eq:minz1}
\end{eqnarray}
where ${\bf{g}}_2^{(i)} = {{\bf{B}}^{\mathsf H}}({\bf{B}}{{\bf{\hat z}}^{(i)}} + {\bf{A}}{{\bf{\hat x}}^{(i + 1)}} - {\bf{y}} + \frac{1}{\rho }{{\bf{s}}^{(i)}})$, and ${{\tau _2}}$ is a positive proximal parameter. After some mathematical manipulations, a more compact form of \eqref{eq:minz1} can be further written as a single measurement vector problem, which is
\begin{eqnarray}
	\mathop {\min }\limits_{\bf{\hat z}} {\lambda _2}\sum\limits_m^M {{w_m}{{\left\| {{{\bf{\hat z}}_m}} \right\|}_1}}  + \frac{1}{2}\left\| {{\bf{\hat z}} - {{\bf{d}}^{(i)}}} \right\|_2^2,
	\label{eq:minz2}
\end{eqnarray}
where ${\lambda _2} = {\tau _2}/\rho $, ${{\bf{d}}^{(i)}} = {{\bf{\hat z}}^{(i)}} - {\tau _2}{\bf{g}}_2^{(i)}$.
The minimization \eqref{eq:minz2} has a closed-form solution \cite{donoho1995noising,chambolle1998nonlinear}, which is given by
\begin{eqnarray}
{[{{\bf{\hat z}}_m^{(i+1)}}]_l} = \left\{ {\begin{array}{*{20}{c}}
	{{{[{{\bf{d}}^{(i)}}]}_{{l_{\rm{s}}}}} - {\lambda _2}{w_m},}&{{{[{{\bf{d}}^{(i)}}]}_{{l_{\rm{s}}}}} > {\lambda _2}{w_m},}\\
	{{{[{{\bf{d}}^{(i)}}]}_{{l_{\rm{s}}}}} + {\lambda _2}{w_m},}&{{{[{{\bf{d}}^{(i)}}]}_{{l_{\rm{s}}}}} <  - {\lambda _2}{w_m},}\\
	{0,}&{{\rm{otherwise,}}}
	\end{array}} \right.
\end{eqnarray}
where ${l_{\rm{s}}} = \sum\nolimits_{j = 1}^{m - 1} {{L_j}}  + l, m = 1,2, \ldots ,M$, and $l = 1,2, \ldots ,{L_m}$.


\vspace{-0.3cm}
\subsection{Deep ADMM Unfolding Architecture}
\begin{figure}[t]
		\hspace{-1cm}
	{\begin{minipage}[h]{0.6\textwidth}
			\centering
			\epsfig{figure=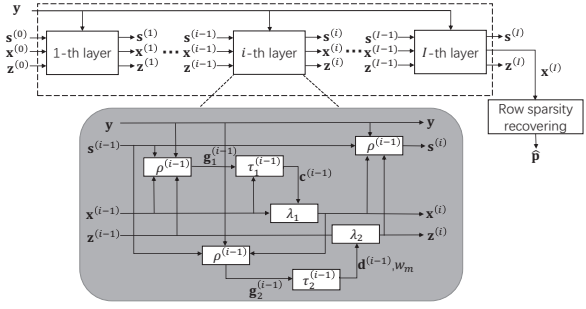,width=0.72\columnwidth}
		\end{minipage}
	}
	\hfill
	\caption{ Signal-flow diagram of DAUN with learnable variables $\rho^{(i)}$, $\{ w_m\} _{m = 1}^M$, ${\tau _1^{(i)}}$, and ${\tau _2^{(i)}}$.    }
	\vspace{-.5cm}
	\label{Fig:2}
\end{figure}

Theoretically, given the proper parameters $\rho$, $\{ {w_m}\} _{m = 1}^M$, $\tau_1$ and $\tau_2$, the ADMM is guaranteed to converge to the optimal solution \cite{lu2011fast}. However, the selections of above parameters are quite heuristic, which requires cumbersome tuning in practice. On the other hand, the parameters obtained from manual tuning are only on the basis of very limited data set. As such, the resulting iterative performance of ADMM is not robust to any upcoming signals generated with arbitrary user position, SNR, number of BS's antennas and etc.
To circumvent such hurdles, we propose to unfold the ADMM iterations from \eqref{eq:minX} to \eqref{eq:minS} into the model-driven deep learning network and, thereby, the unknown parameters can be learned from the DAUN.
Different from the black-box data-driven neural network, the DAUN unfolds the ADMM iteration into a trainable framework. As such, the interpretability of the ADMM update translates directly to the proposed DAUN.
In DAUN, the parameters $\rho$, $\{ {w_m}\} _{m = 1}^M$, $\tau_1$, $\tau_2$ are untied across the ADMM iterations so that at the $i$-th iteration, the parameters $\rho^{(i)}$, $\{ w_m\} _{m = 1}^M$ $\tau_1^{(i)}$, $\tau_2^{(i)}$ are trainable. Note here we set different paramter $\rho^{(i)}$, $\tau_1^{(i)}$, $\tau_2^{(i)}$ rather than set them fixed ones in each layer for better generality and convergence speed of the network \cite{balatsoukas2019deep}. The signal-flow graph of a $I$-layer DAUN is given in Fig. \ref{Fig:2}, where the
nodes in the graph represent the different operations in the ADMM updates and the directed edges represent the data flows between these operations, the number of total trainable parameters is $3I + M$.

Since a sharp reduction of localization accuracy may occurs at low SNR region, to alleviate such degeneration, we propose to incorporate a SNR-weighted loss function into DAUN, which is given as
\vspace{-0.5em}
\begin{eqnarray}
{\cal L}({\bf{y}},{{{\bf{\hat x}}}^{(i)}},{{{\bf{\hat z}}}^{(i)}}){\rm{ = (1 + }}\sigma {\rm{( - }}\bar \omega {\rm{))}}\left\| {{\bf{y}} - {\bf{A}}{{{\bf{\hat x}}}^{(i)}} + {\bf{B}}{{{\bf{\hat z}}}^{(i)}}} \right\|_2^2,
	\label{eq:snr}
\end{eqnarray}
where ${\bar \omega }$ is the SNR (in dB) associated with the received signal ${\bf{y}}$, and $\sigma (x) = 1/(1 + {e^{ - x}})$ denotes the sigmoid function.

In the model training phase, the received signal ${\bf{y}}$ is independently generated, where the user position is randomly drawn from the target area $\mathcal R$ and the SNR is uniformly distributed from the region of interest.
We adopt mini-batch training with a stochastic gradient descent algorithm for DAUN.  To overcome the problem of vanishing gradient, an incremental training regime is adopted, in which all the variables are sequentially learned from the first layer to the last layer. After the training of the
first $i$ layers, the new $(i+1)$-th layer is appended to the DAUN and the entire network is trained again. The variables $\{ {\rho ^{(t)}}, \{ w_m^{(i)}\} _{m = 1}^M, \tau _1^{(t)},\tau _2^{(t)}\} _{t = 1}^i$ are taken as the initial values in the new training process. This operation can effectively reduce the training complexity, as indicated in \cite{balatsoukas2019deep} and our simulations.

\vspace{-0.3cm}
\subsection{Position Refinement}
Physically, the localization accuracy of DAUN relies on a correct classification of the BSs with active LoS paths. When some BSs have no active LoS paths, a simple selection $\hat k = \mathop {\arg \max }_k {\left\| {[{{\bf{\hat X}}]_{k,:}}} \right\|_2}$ may yield a suboptimal position estimate. This is because the non-zero elements in ${[{\bf{\hat X}}]_{k,:}}$ corresponding to the inactive LoS paths may make ${\left\| {{{[{\bf{\hat X}}]}_{\hat k,:}}} \right\|_2}$ the largest one. For example, we assume that the true user position is associated with ${[{\bf{X}}]_{{k_1},:}}$, i.e., the grid location ${{\bf{\Phi }}_{{k_1}}}$ should be the position estimate. By using the ADMM update, we obtain the estimate ${[{\bf{\hat X}}]_{{k_1},:}} = [1.6,1.5,1.7,0.1]$, where the first three elements in ${[{\bf{\hat X}}]_{{k_1},:}}$ are the active LoS path gain estimates and the last element in ${[{\bf{\hat X}}]_{{k_1},:}}$ can be seen as a noise term corresponding to the inactive LoS path. Meanwhile, for another index $k_2$ around $k_1$, we may obtain the estimate ${[{\bf{\hat X}}]_{{k_2},:}} = [1.55,1.45,1.7,0.25]$. We can find that by performing $\hat k = \mathop {\arg \max }_k {\left\| {[{{\bf{\hat X}}]_{k,:}}} \right\|_2}$, the wrong index $k_2$ will be returned instead of $k_1$. To avoid such mismatch, we propose to cluster the active elements of ${{{[{\bf{\hat X}}]}_{k,:}}}$ into a set ${\mathcal{S}}_{k,1}$ such that the BSs with possible active LoS paths can be coarsely selected. Then, we obtain the position estimate ${\hat k}$ within the set ${\mathcal{F}}$ associated with the BSs with active LoS paths. As such, the position estimate $\hat k$ is determined without the effect of inactive LoS components. The descriptions of above position refinement process is detailed in Algorithm 1.

$\textit{Remark 2:}$  The performance of direct localization depends heavily on the angular resolution of each BS array such that the grid number $K$ and antenna number $N_m$ are expected to be as large as possible to achieve a better localization accuracy. To solve problem \eqref{eq:direct}, the complexity of MOSEK solver in \cite{garcia2017direct} is ${\mathcal{O}}({(KM + \sum\nolimits_{m = 1}^M {{L_m}} )^{3.5}}\sum\nolimits_{m = 1}^M {{N_m}} )$, which is computationally heavy for real-time implementations, especially with mmWave massive arrays. Moreover, the direct position determination in \cite{weiss2004direct} requires high-dimension singular value decomposition and exhaustive gird search, which also has a huge complexity. Appealingly, the proposed ADMM is a first order primal-dual algorithm, in which the calculations of  \eqref{eq:minX}, \eqref{eq:minz} and \eqref{eq:minS} only involve linear operations, e.g., simple matrix multiplications/additions. In each iteration of the ADMM, the complexity is only ${\mathcal{O}}(({(MK)^2} + {(\sum\nolimits_{m = 1}^M {{L_m}} )^2})\sum\nolimits_{m = 1}^M {{N_m}} )$.
Moreover, the DAUN is trained in an offline manner and the complexity of the position refinement is only ${\mathcal O}(KM)$. As such, the complexity of DAUN is dominated by the ADMM update, which is quite low, as validated in Section \ref{sec:simu}.

\begin{algorithm} [t!]
\small
	\caption{Position Refinement}
	\label{alg:ALG1}
	\begin{algorithmic}[1]
		\State Initialization: ${\chi _{k,1}} = \max ({[{\bf{\hat X}}]_{k,:}}),{\chi _{k,2}} = 0$, $\forall k$
		\While  {${\mathcal{S}}_{k,1}$ and ${\mathcal{S}}_{k,2}$ are changing}
		\State  ${\mathcal{S}}_{k,1} = \emptyset $ and ${\mathcal{S}}_{k,2} = \emptyset $
		\For {$i=1$ to $M$}
		\If {$|{[{\bf{\hat X}}]_{k,i}} - {\chi _{k,1}}| > |{[{\bf{\hat X}}]_{k,i}} - {\chi _{k,2}}|$} {${{\mathcal{S}}_{k,1}} = {{\mathcal{S}}_{k,1}} \cup \{ i\} $}
		\Else~ {${{\mathcal{S}}_{k,2}} = {{\mathcal{S}}_{k,2}} \cup \{ i\} $}
		\EndIf
		\State ${\chi _{k,1}} = \sum\nolimits_{m \in {{\mathcal{S}}_{k,1}}} {\frac{{{{[{\bf{\hat X}}]}_{k,m}}}}{{|{{\mathcal{S}}_{k,1}}|}}} ,{\chi _{k,2}} = \sum\nolimits_{m \in {{\mathcal{S}}_{k,2}}} {\frac{{{{[{\bf{\hat X}}]}_{k,m}}}}{{|{{\mathcal{S}}_{k,2}}|}}} $
		\EndFor
		\EndWhile
		\State Finding those indexes $\tilde k$ such that ${\rm{|}}{{\mathcal{S}}_{\tilde k, 1}}| \ge 3$ and collect them into a set ${\mathcal{F}}$
		\State $\hat k = \mathop {\arg \max }\limits_{\tilde k} {\left\| {{{[{\bf{\hat X}}]}_{\tilde k,{{\mathcal{S}}_{\tilde k,1}}}}} \right\|_2}, \tilde k \in {\mathcal{F}}$
	\end{algorithmic}
\end{algorithm}

\vspace{-0.5cm}
\section{Numerical Results} \label{sec:simu}
In this section, unless otherwise stated, all experiments were run using the following parameters. We considered a situation that the origin of the coordinate system is in the middle of the area and there were $M=4$ BSs locating at [-50m,-50m], [-50m, 50m], [50m, 50m], [50m, -50m], respectively. All the BSs were equipped with uniform linear arrays (ULAs) and the inter antenna spacing was equal to half wavelength. The number of antennas at each BS was ${N_m} = 50,m=1,2,3,4$. The user was located within the target area $\mathcal{L}$ with size $40 \times 40$ m. To exploit the optimal localization performance, the numbers of grid locations and discrete angles were $K=30 \times 30 = 900$ and $L_m = 100, \forall m$, respectively. The mmWave carrier frequency was 30 GHz and the channels were generated under the 3GPP Urban Macro Scenario \cite{haneda20165g}.
The DAUN was implemented in Python using the Pytorch library \cite{paszke2017automatic}. We trained the network using 700 samples per new layer with the batch size of 7. The Adam optimizer was applied with the learning rate of 0.05 and reduced it to 0.01 when the total layers was larger than 5. The validation sets contained 200 samples, and every training process was terminated when the validation loss was not decreasing. Note that the Pytorch can only tackle the real-value deep learning problems. Hence, in the model training phase, the direct localization problem \eqref{eq:direct} were transformed into its equivalent real form. All the results were obtained by running $1000$ Monte Carlo experiments.

\begin{figure}[t!]
	{\begin{minipage}[h]{0.58\textwidth}
			\centering
			\epsfig{figure=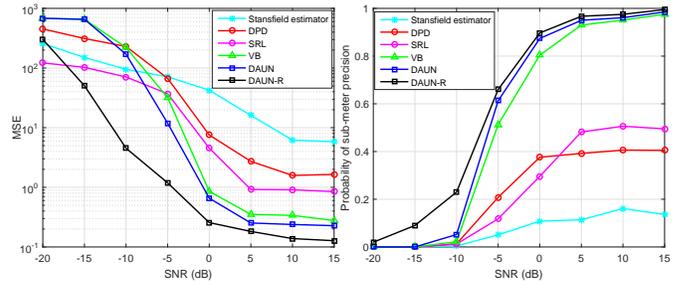,width=1.1\columnwidth}
		\end{minipage}
	}
	\hfill
	\caption{Performance comparisons of different localization methods. (a) The MSE of position estimates as a function of the SNR; (b)The probability of sub-meter precision as a function of the SNR. }
	\vspace{-.5cm}
\label{Fig_3}
\end{figure}
	
First, the performance comparisons were shown in Fig. \ref{Fig_3}, where the proposed DAUN  and its position refinement variant (DAUN-R) were compared with direct position determination (DPD) \cite{weiss2004direct}, Stansfield estimator \cite{gavish1992performance}, sparse reconstruction localization (SRL) \cite{malioutov2005sparse}, and modified variational Bayes (VB) \cite{tzikas2008variational}. Besides using the mean square error (MSE) ${\rm MSE = }{\mathbb{E}}[\left\| {{\bf{\hat p}} - {\bf{p}}} \right\|_2^2]$, the probability of sub-meter precision
was also adopted. This was because for many applications, e.g., autonomous driving, Internet-of-Things and rescue operations, the sub-meter localization precision is requisite \cite{zhao2020beamspace,shahmansoori2017position}.
We found that DAUN-R achieved lower MSE and higher sub-meter accuracy over a wide SNR range, as compared with the other techniques. With the increasing of SNR, the probability of sub-meter precision of DAUN and DAUN-R were both approaching to unit.

\begin{figure}[t!]
		\hspace{-0.5cm}
	{\begin{minipage}[h]{0.5\textwidth}
			\centering
			\epsfig{figure=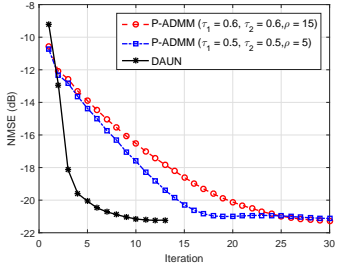,width=0.55\columnwidth}
		\end{minipage}
	}
	\hfill
	\caption{NMSE of DAUN and P-ADMM as a function of iteration. }
	\vspace{-.3cm}
	\label{Fig_4}
\end{figure}

Moreover, the proposed DAUN solver had better iterative performance, as compared with its pure ADMM (P-ADMM) counterpart. To illustrate this superiority, Fig. \ref{Fig_4} compared the convergence speed of DAUN and P-ADMM, where the normalized mean squared error ${\rm{NMSE = \mathbb{E}[}}\left\| {{\bf{y}} - {\bf{A \hat x}} - {\bf{B \hat z}}} \right\|_2^2{\rm{]/\mathbb{E}[}}\left\| {\bf{y}} \right\|_2^2{\rm{]}}$ varied with respect to the iterations and the parameters of P-ADMM were selected through multiple experiments to guarantee the convergence. The figure showed that both the DAUN and P-ADMM converged monotonically with respect to the iteration. We observed that the convergence speed of DAUN was much faster than that of P-ADMM. To further illustrate low complexity of DAUN, the average computational time of different localization solvers were given in Table \ref{tab_1}, where all the programs were running on Inter Core i7-8700K @4.2 GHz$\times$12 cores. It showed that the computational burden of DAUN is greatly reduced, thanks to the low-complexity ADMM and the fine parameter tuning of unfolding network.

\begin{table}[!t]
\caption{Average Computational Time of Different Direct Localization Solvers (DAUN, P-ADMM ($\tau_1 = 0.5$, $\tau_2 = 0.5$, $\rho = 5$), MOSEK \cite{garcia2017direct}, Majorization-minimization (MM) \cite{zheng2020joint}, Modified Variational Bayes (VB) \cite{tzikas2008variational}, and DPD \cite{weiss2004direct}).}
\centering
\begin{tabular}{c c c c c c c} 
		\hline
		\hline
		 & DAUN & P-ADMM & MOSEK & MM & VB & DPD \\
		\hline
		Time (sec) & 0.017 & 0.102 & 4.27 & 3.81 & 1.36 & 2.15 \\	
		\hline	
		\hline
\end{tabular}
\vspace{-0.2cm}
\label{tab_1}
\end{table}

\section{Conclusions}
In this paper, we have proposed the DAUN for fast direct localization in mmWave MIMO systems. First, the direct localization has been formulated as a joint $l_1$-$l_{2,1}$ norm  sparse recovery problem. Then, we have developed the ADMM solver to tackle this problem. Since the convergence of ADMM heavily relied on fine parameter tuning, we have adapted the deep unfolding into the ADMM iteration so as to learn the parameters in a network training manner.  Finally, a position refinement algorithm has been designed to further improve the localization accuracy of DAUN. To the best of our knowledge, the proposed DAUN is the fastest algorithm to achieve the direct localization for mmWave massive MIMO systems, which is of great importance from implementation perspective.

\balance
\small
\bibliographystyle{IEEEbib}
\bibliography{yourbibliography}

\begin{thebibliography}{10}

\bibitem{xiao2017millimeter}
M.~Xiao, S.~Mumtaz, Y.~Huang, L.~Dai, Y.~Li, M.~Matthaiou, G.~K. Karagiannidis,
  E.~Bj{\"o}rnson, K.~Yang, C.-L. I, and A.~Ghosh,
\newblock ``Millimeter wave communications for future mobile networks,''
\newblock {\em IEEE J. Sel. Areas Commun.}, vol. 35, no. 9, pp. 1909--1935,
  Sep. 2017.

\bibitem{shahmansoori2017position}
A.~Shahmansoori, G.~E. Garcia, G.~Destino, G.~Seco-Granados, and H.~Wymeersch,
\newblock ``Position and orientation estimation through millimeter-wave {MIMO}
  in {5G} systems,''
\newblock {\em IEEE Trans. Wireless Commun.}, vol. 17, no. 3, pp. 1822--1835,
  Mar. 2017.

\bibitem{pan2022efficient}
M.~Pan, P.~Liu, S.~Liu, W.~Qi, Y.~Huang, X.~You, X.~Jia, and X.~Li,
\newblock ``Efficient joint {DOA} and {TOA} estimation for indoor positioning
  with {5G} picocell base stations,''
\newblock {\em IEEE Trans. Instrum. Meas.}, vol. 71, no. 8005219, Aug. 2022.

\bibitem{hu2014esprit}
A.~Hu, T.~Lv, H.~Gao, Z.~Zhang, and S.~Yang,
\newblock ``An {ESPRIT-based} approach for {2-D} localization of incoherently
  distributed sources in massive {MIMO} systems,''
\newblock {\em IEEE J. Sel. Topics Signal Process.}, vol. 8, no. 5, pp.
  996--1011, Oct. 2014.

\bibitem{wang2015asymptotically}
Y.~Wang and K.~Ho,
\newblock ``An asymptotically efficient estimator in closed-form for {3-D AOA}
  localization using a sensor network,''
\newblock {\em IEEE Trans. Wireless Commun.}, vol. 14, no. 12, pp. 6524--6535,
  Dec. 2015.

\bibitem{yang2021model}
J.~Yang, S.~Jin, C.-K. Wen, J.~Guo, M.~Matthaiou, and B.~Gao,
\newblock ``Model-based learning network for {3-D} localization in mmwave
  communications,''
\newblock {\em IEEE Trans. Wireless Commun.}, vol. 20, no. 8, pp. 5449--5466,
  Aug. 2021.

\bibitem{zhao2020beamspace}
H.~Zhao, N.~Zhang, and Y.~Shen,
\newblock ``Beamspace direct localization for large-scale antenna array
  systems,''
\newblock {\em IEEE Trans. Signal Process.}, vol. 68, pp. 3529--3544, May.
  2020.

\bibitem{weiss2004direct}
A.~J. Weiss,
\newblock ``Direct position determination of narrowband radio frequency
  transmitters,''
\newblock {\em IEEE Signal Process. Lett.}, vol. 11, no. 5, pp. 513--516, May
  2004.

\bibitem{garcia2017direct}
N.~Garcia, H.~Wymeersch, E.~G. Larsson, A.~M. Haimovich, and M.~Coulon,
\newblock ``Direct localization for massive {MIMO},''
\newblock {\em IEEE Trans. Signal Process.}, vol. 65, no. 10, pp. 2475--2487,
  May. 2017.

\bibitem{lian2019user}
L.~Lian, A.~Liu, and V.~K. Lau,
\newblock ``User location tracking in massive {MIMO} systems via dynamic
  variational {Bayesian} inference,''
\newblock {\em IEEE Trans. Signal Process.}, vol. 67, no. 21, pp. 5628--5642,
  Nov. 2019.

\bibitem{zheng2020joint}
X.~Zheng, A.~Liu, and V.~Lau,
\newblock ``Joint channel and location estimation of massive {MIMO} system with
  phase noise,''
\newblock {\em IEEE Trans. Signal Process.}, vol. 68, pp. 2598--2612, Apr.
  2020.

\bibitem{li2022self}
J.~Li, P.~Li, P.~Li, L.~Tang, X.~Zhang, and Q.~Wu,
\newblock ``Self-position awareness based on cascade direct localization over
  multiple source data,''
\newblock {\em IEEE Trans. Intell. Transp. Syst.}, early access, May 2022.

\bibitem{chambolle1998nonlinear}
A.~Chambolle, R.~A. DeVore, N.~Lee, and B.~J. Lucier,
\newblock ``Nonlinear wavelet image processing: variational problems,
  compression, and noise removal through wavelet shrinkage,''
\newblock {\em IEEE Trans. Image Process.}, vol. 7, no. 3, pp. 319--335, Mar.
  1998.

\bibitem{rangan2019convergence}
S.~Rangan, P.~Schniter, A.~K. Fletcher, and S.~Sarkar,
\newblock ``On the convergence of approximate message passing with arbitrary
  matrices,''
\newblock {\em IEEE Trans. Inf. Theory.}, vol. 65, no. 9, pp. 5339--5351, Sep.
  2019.

\bibitem{tzikas2008variational}
D.~G. Tzikas, A.~C. Likas, and N.~P. Galatsanos,
\newblock ``The variational approximation for {Bayesian} inference,''
\newblock {\em IEEE Signal Process. Mag.}, vol. 25, no. 6, pp. 131--146, Nov.
  2008.

\bibitem{donoho1995noising}
D.~L. Donoho,
\newblock ``De-noising by soft-thresholding,''
\newblock {\em IEEE Trans. Inf. Theory}, vol. 41, no. 3, pp. 613--627, May
  1995.

\bibitem{lu2011fast}
H.~Lu, X.~Long, and J.~Lv,
\newblock ``A fast algorithm for recovery of jointly sparse vectors based on
  the alternating direction methods,''
\newblock in {\em Proc. 14th Int. Conf. Artificial Intelligence and
  Statistics}, 2011, pp. 461--469.

\bibitem{balatsoukas2019deep}
A.~B. Stimming and C.~Studer,
\newblock ``Deep unfolding for communications systems: A survey and some new
  directions,''
\newblock in {\em IEEE Int. Workshop on Signal Process. Systems (SiPS)}. IEEE,
  Oct. 2019, pp. 266--271.

\bibitem{haneda20165g}
K.~Haneda et~al.,
\newblock ``{5G} {3GPP-like} channel models for outdoor urban microcellular and
  macrocellular environments,''
\newblock in {\em Proc. IEEE 83rd Veh. Technol. Conf. (VTC Spring)}, 2016, pp.
  1--7.

\bibitem{paszke2017automatic}
A.~Paszke et~al.,
\newblock ``Automatic differentiation in {PyTorch},''
\newblock in {\em Proc. 31st Conf. Neural Inf. Process. Syst.}, 2017, pp. 1--4.

\bibitem{gavish1992performance}
M.~Gavish and A.~J. Weiss,
\newblock ``Performance analysis of bearing-only target location algorithms,''
\newblock {\em IEEE Trans. Aerosp. Electron. Syst.}, vol. 28, no. 3, pp.
  817--828, Jul. 1992.

\bibitem{malioutov2005sparse}
D.~Malioutov, M.~Cetin, and A.~S. Willsky,
\newblock ``A sparse signal reconstruction perspective for source localization
  with sensor arrays,''
\newblock {\em IEEE Trans. Signal Proces}, vol. 53, no. 8, pp. 3010--3022, Aug.
  2005.

\end{thebibliography}

\end{document}